\documentclass[aps,prx,floatfix,twocolumn,showpacs]{revtex4}

\usepackage{graphicx}
\usepackage{epsfig}
\usepackage[usenames]{xcolor}

\begin{document}

\title{Collapsing Schr\"odinger Cats in the Density Matrix Renormalization Group}

\author{Hong-Chen Jiang and Leon Balents} 
\affiliation{Kavli Institute for Theoretical Physics, University of California, Santa Barbara, CA, 93106}

\date{\today}

\begin{abstract}%
In this paper, we propose a modified Density Matrix Renormalization Group (DMRG) algorithm to preferentially select minimum entropy states (minimally entangled states) in finite systems with asymptotic ground state degeneracy.  The algorithm adds a ``quench'' process to the conventional DMRG method, which mimics the decoherence of physical systems, and collapses non-locally entangled states such as Schr\"odinger cats.  We show that the method works for representative models with ground state degeneracy arising from either topological order or spontaneous discrete symmetry breaking.  In the minimal entropy states thus obtained, properties associated with thermodynamic limit, such as topological entanglement entropy and magnetic order parameters, can be obtained directly and efficiently.
\end{abstract}

\pacs{75.10.Jm, 75.40.Mg, 75.50.Ee}

\maketitle

\section{Introduction}%

Numerical methods such as exact diagonalization, quantum Monte Carlo, and Density Matrix Renormalization Group (DMRG)\cite{White1992DMRG,White1993DMRG,DMRG_RMP,Schollwock2011DMRG} have become powerful numerically exact techniques for studying substantial finite quantum systems.  However, in most applications, actual physical systems are enormously larger than the limits of these methods, and can be considered in the thermodynamic limit, i.e. infinite.  An important distinct between even large finite systems and infinite ones can be made in terms of non-local entanglement.  In a large finite system, non-local entanglement arises in the presence of spontaneous symmetry breaking and/or intrinsic topological order.  In the former case, typically spontaneous symmetry breaking is avoided for finite systems even at zero temperature by quantum tunneling, which results in an {\sl absolute} system ground state (we will refer to the unique ground state of a finite system as the absolute ground state) which is a ``Schr\"odinger cat'' state, a superposition of macroscopically distinct ordered states.  For example, in a quantum Ising ferromagnet, the true ground state of a finite system is an equal weight superposition of the two opposite magnetization states\cite{SachdevPhysToday}.  The lowest excited state in that case is an orthogonal linear combination of the magnetization states.  In the thermodynamic limit these two states become degenerate, and spontaneous symmetry breaking occurs in real systems in response to arbitrarily weak perturbations or decoherence which ``collapses'' the Schr\"odinger cat state in favor of a single macroscopic ground state which breaks the symmetry.  

Because such decoherence processes are inevitable in experimental systems, typically one is most interested in properties of a single macroscopic ground state.  In this respect, the non-local entanglement of the absolute ground state of a finite system is a nuisance.  It is therefore desirable to avoid it in a numerical simulation.  It is preferable to obtain, instead of the absolute ground state, a {\em Minimally Entangled State} or {\em Minimal Entropy State} (MES)\cite{Zhang2012EE}, which is a linear combination of the degenerate ground states which is chosen to minimize non-local entanglement.  For the case of spontaneous symmetry breaking, a MES corresponds to the state physically selected by decoherence.   In this paper, we propose an improved DMRG procedure which combines the standard DMRG algorithm and a new element which we denote a ``quantum quench'', which directly obtains a MES.  We demonstrate by examples that the improved DMRG procedure performs well for gapped systems, including those exhibiting topological order and global discrete symmetry breaking.  Furthermore, it also works for gapless systems in which extensive ground state degeneracy arises from redundant degrees of freedom.   Using the improved procedure to obtain a MES, we can directly study the thermodynamic limit and obtain, for example, highly accurate values for the order parameter in symmetry broken states, and for the topological entanglement entropy of topological phases, using significantly less computational effort than is necessary to extract those quantities from the conventional DMRG.

\section{Improved DMRG procedure}
\label{Sec:ImprovedDMRG}

The density-matrix renormalization group (DMRG) is a numerical technique for systematically improving variationally states of quantum Hamiltonians.   Since its introduction in 1992\cite{White1992DMRG}, DMRG has quickly achieved the status of a highly reliable and precise numerical method for studying one-dimensional\cite{White1993DMRG,DMRG_RMP,Jiang2010S2Chain,Stoudenmire2011Majorana} and quasi-one-dimensional strongly correlated quantum systems\cite{Jiang2008Kagome,Jiang2009S1,Stoudenmire2011,
White2011Kagome,Jiang2011SJ1J2,Depenbrock2012Kagome,Jiang2012TEE,Cincio2013TO}. Its main advantages are the ability to treat large systems with high accuracy at zero temperature and the absence of the negative sign problem that plagues the powerful quantum Monte Carlo.

\begin{figure}
\centerline{
    \includegraphics[height=2.4in,width=3.0in] {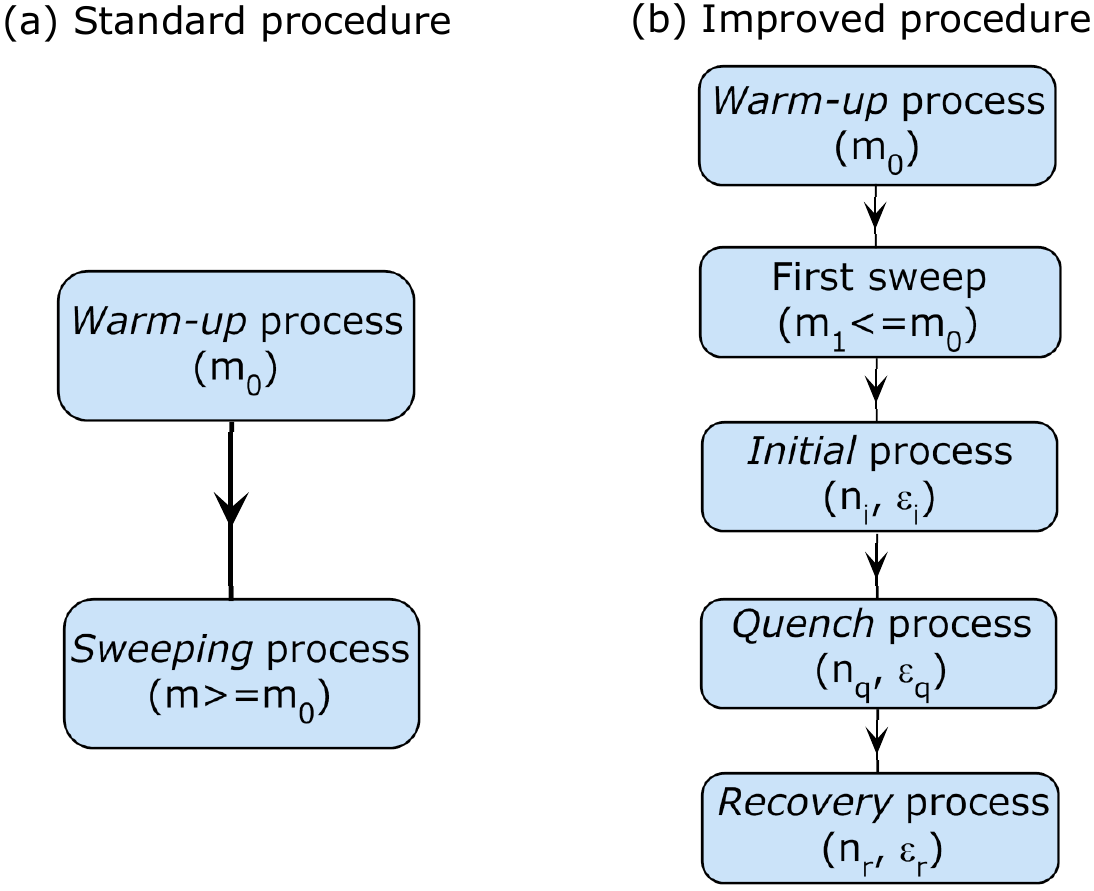}
    }
\caption{(Color online) Flow chart of the (a) standard DMRG procedure and (b)
improved DMRG procedure (See main text and Table.\ref{Tab:ImprovedDMRG} for detail). $m_0$ and $m_1$ denote the minimal number of states kept in the \emph{warm-up} process and \emph{sweeping} processes. ($n_i$,$\varepsilon_i$), ($n_q$,$\varepsilon_q$) and ($n_r$,$\varepsilon_r$) are the number of sweeps and total truncation errors used in the \emph{initial}, \emph{quench}, and \emph{recovery} processes respectively.}
\label{Fig:DMRGPorcedure}
\end{figure}

The standard DMRG algorithm has two steps, known as the \emph{warm-up} process (infinite-system algorithm) and \emph{sweeping} process (finite-system algorithm), as shown in Fig.\ref{Fig:DMRGPorcedure}(a).  In the \emph{warm-up} process, a wavefunction of an increasingly large system is iteratively constructed, until the desired length is reached.  Then, more accurate and converged results are achieved by the \emph{sweeping} process.  Integral to the DMRG is an efficient and systematic truncation procedure for keeping a small number of important states, taken as those with the largest weight in the Schmidt decomposition of the reduced density matrix. Typically the number of Schmidt states kept $m$ is chosen in order to achieve a desired truncation error $\varepsilon$ for the reduced density matrix, and also constrained to be greater than or equal to some minimum value $m_0$ (see Supplemental Information, Fig.\ref{Fig:DMRGPorcedure} and Ref.\cite{White1992DMRG,White1993DMRG,DMRG_RMP,Schollwock2011DMRG} for more details). In practical applications, one often starts building up the system with a relatively small number of states $m_0$ in the \emph{warm-up} process, and increasing it to $m\geq m_0$ while running through the sweeps until converged results are achieved.

The standard DMRG procedure is extremely useful and applicable to many problems\cite{Jiang2008Kagome,Jiang2009S1,Stoudenmire2011,
White2011Kagome,Jiang2011SJ1J2,Depenbrock2012Kagome,Jiang2012TEE,Cincio2013TO}. However, in the presence of ground state degeneracy, the complications associated with non-local entanglement and Schr\"odinger cat states described in the introduction arise.  Examples will be discussed in later sections.   Using standard DMRG, these complications can be surmounted for very long cylinders at exponentially large computational cost (see for example \cite{Jiang2012TEE,Jiang2013TEEAccuracy}).   Here, we instead propose an improved procedure which modifies the standard DMRG algorithm specifically to obtain the desired MES without such high computational cost, and on systems of modest length.  The improved procedure is shown in Fig.\ref{Fig:DMRGPorcedure}(b).    Our empirical studies demonstrate that the improved procedure works for gapped systems with spontaneous symmetry breaking or topological order, as well as for an example with an extensive ground state degeneracy coming from redundant degrees of freedom (the 1D Kitaev chain in Sec.\ref{Sec:MajoranaChain}).

\begin{table}[t]
\caption{Improved DMRG procedure, as shown in Fig.\ref{Fig:DMRGPorcedure}(b).} \label{Tab:ImprovedDMRG}
\begin{tabular}{l p{8cm}} 
\hline\hline
1. & Use the standard \emph{warm-up} process to build up the system to the desired length,
using a moderate lower bound on the number of states $m_0$. \\ 
2. & Finish the warm-up with a first sweep, keeping a number of states with a smaller lower bound $m_1 \leq m_0$. \\%
3. & In the \emph{initial} process, perform a set of $n_{i}$ sweeps with a fixed total truncation error $\varepsilon_{i}$.   From this stage onward, the lower bound on the number of states is set to unity, i.e. $m\geq 1$. \\%
4. & At the beginning of the \emph{quench} process, suddenly increase the truncation error from $\varepsilon_{i}$ to a larger value $\varepsilon_{q}$, and perform another $n_q$ sweeps.\\%
5. & Finally, in the \emph{recovery} process, change the truncation error back to a smaller value $\varepsilon_{r}$, and carry out a sufficient number, $n_r$, of sweeps, so that the final results are  converged.\\%
\hline\hline%
\end{tabular}
\end{table}

The detailed scheme of the improved procedure is summarized in Fig.\ref{Fig:DMRGPorcedure}(b) and Table.\ref{Tab:ImprovedDMRG}. It consists of four stages: {\em warm-up} (expanded to include a first sweep), {\em initial} , {\em quench}, and {\em recovery}, which we now describe. The improved procedure begins with the standard \emph{warm-up} process to build up the system to the desired length, using a moderate lower bound $m_0$ on the minimum number of states. In practice, $m_0$ should not be taken so small that convergence becomes slow, but could in principle be as small a $1$. To speed the later convergence, a first sweep is then made.  However, contrary to the usual DMRG procedure, we {\em decrease} the lower bound on the number of states during this sweep to a new value $m_1 \leq m_0$. This speeds the later convergence to the MES, as the entanglement is reduced by this process.

Following the \emph{warm-up} process, we carry out a set of $n_i$ sweeps at moderate fixed total truncation error $\varepsilon_i$, with the lower bound on the number of states $m$ taken to the minimum value of unity (indeed we keep this minimum value at $m\geq 1$ for all the remaining processes).  The actual kept number of states is controlled by the total truncation error $\varepsilon_{i}$ and automatically determined by the simulation itself.  We refer to these sweeps (typically $n_i \approx 10$) as the \emph{initial} process.   Like the first sweep, the \emph{initial} process promotes faster convergence to the MES, compared with the standard procedure, because the actual number of states needed for the MES can be smaller than $m_0$. Examples are shown in Fig.\ref{FigS:TCMM0MaveMES}, in which the averaged number of states $m_{ave}$ required to obtain the MES for the Toric-Code model is plotted.    

The above two stages already favor the MES more than in the standard DMRG procedure.  However, in some cases, this is not sufficient.  In the third stage we introduce the \emph{quench} process, which removes the non-local entanglement by mimicking the decoherence phenomena of physical systems.  In particular, starting with the state $|\psi_i\rangle$ prepared by the \emph{initial} process, we suddenly {\em increase} the truncation error to a new value $\varepsilon_{q}$, and a subsequent state $|\psi_q\rangle$ is obtained by performing another $n_q$ sweeps with this value.  We use the term ``quench'' because the change in truncation error is sudden, though the nomenclature is perhaps not ideal since an {\em increase} in $\varepsilon$ corresponds more closely to an increase in temperature rather than a decrease, as in a usual quench. Typically, the new truncation error $\varepsilon_{q}$ may be several orders of magnitude larger than $\varepsilon_{i}$, so that the number of states associated with $\varepsilon_{q}$ becomes smaller than that in the \emph{initial} process. Consequently, the the entanglement spectrum of the reduced density matrix is significantly altered by the \emph{quench} process, as shown in the inset of Fig.\ref{Fig:TCMHx03Hz00MS}(b) for the Toric-Code model.

After the \emph{quench} process, non-local entanglement has been removed, and we seek to restore the accuracy of the MES wavefunction.  This is done by the \emph{recovery} process, in which the truncation error is decreased from $\varepsilon_q$ to the final value $\varepsilon_r$, which can be smaller than $\varepsilon_i$, i.e., $\varepsilon_r\leq \varepsilon_i$.  A sufficient number of sweeps $n_r$ is performed at this truncation error to achieve a final converged result, i.e., $|\psi_r\rangle$, the MES of the system.

In practical simulations, we found that the parameters characterizing the \emph{initial} and \emph{recovery} stages could be chosen similarly for all models. We took a relative small total truncation error $\varepsilon_i$ (e.g., $10^{-5}$) in the \emph{initial} process, and a (much) smaller value $\varepsilon_r$ (e.g., $10^{-6}$ or smaller) in the \emph{recovery} process, to achieve fast and accurate convergence to the MES.   The typical number of sweeps in the DMRG simulation for the models studied in this paper was $n_i\approx 10$ and $n_r\approx 20$, with slight variations as indicated in the text.  However, the choice of parameters in the \emph{quench} process needed for optimal performance was more dependent on model and system width. One rule of thumb is that the truncation error $\varepsilon_q$ should be larger for thinner systems with smaller entanglement entropy, and conversely smaller for wider systems with larger entanglement entropy. Examples of the detailed truncation error values $\varepsilon_q$ needed to obtain the MES for the Toric-Code model are given in Fig.\ref{FigS:TCMNErrQMES}  in Appendix \ref{appendixTCM}.

\section{Toric-Code Model}
\label{Sec:ToricCodeModel}

We now begin with the well-known Toric-Code model (TCM)\cite{Kitaev2003} (See Appendix \ref{appendixTCM} for notational details). Without magnetic field, i.e., $h_x=h_z=0$, the pure TCM is exactly solvable\cite{Kitaev2003}, and has a $Z_2$ topological ordered ground state. After turning on the magnetic field, the model is no longer exactly solvable. However, previous studies\cite{Trebst2007,Vidal2008,Tupitsyn2008} show that the $Z_2$ topological phase remains stable and robust until the magnetic fields are large enough that the system undergoes a transition from the topological phase to the trivial one. Specifically, such a phase transition takes place at the critical magnetic field $h_x^c\approx 0.328$ when $h_z=0$, while $h_x^c\approx 0.34$ along the symmetric line with $h_z=h_x$.

\begin{figure}
\centerline{
    \includegraphics[height=4.4in,width=3.4in] {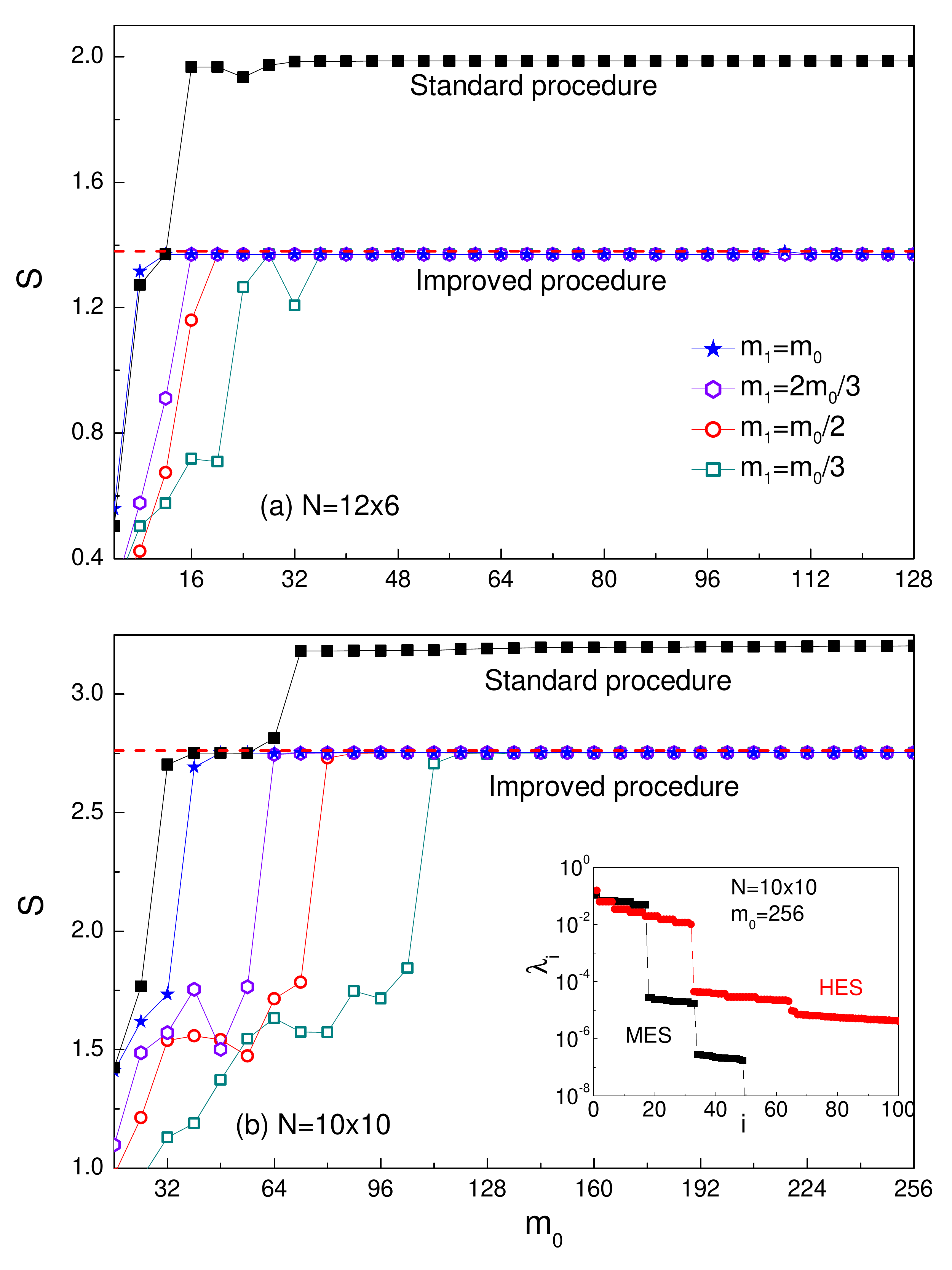}
    }
\caption{(Color online) The von Neumann entanglement entropy $S$ versus
the number of states $m_0$ with different $m_1$ for the Toric-Code model on a cylinder at $h_x=0.3$ and $h_z=0.0$, for (a) $N=12\times 6$ and (b) $N=10\times 10$. Red dashed lines denote the results in the long-cylinder limit $L_x=\infty$, black squares denote the results obtained using the standard procedure, and the rest are  obtained using the improved procedure. Inset in (b): Entanglement spectrum \{$\lambda_i$\} for the TCM with $N=10\times 10$ and $m_0=256$, obtained using the standard procedure with HES (red circles), and the improved procedure with MES (black squares), respectively. Here other parameters are $\varepsilon_r=10^{-6}$, $n_q=4$, $\varepsilon_q=5\times 10^{-3}$ for $L_y=6$, and $\varepsilon_q=10^{-3}$ for $L_y=10$.} \label{Fig:TCMHx03Hz00MS}
\end{figure}

In the $Z_2$ topological phase, there are two-fold quasi-degenerate ground states if we put the system on a cylinder. Interestingly, Jiang et al\cite{Jiang2012TEE} showed that the conventional DMRG algorithm will naturally find a MES in the long cylinder limit, i.e., $L_x=\infty$. Therefore, an accurate topological entanglement entropy (TEE) can be extrapolated by fitting $S(L_y)=\alpha L_y-\gamma$, even very close to the phase transition point.   However, we note that this result within the conventional algorithm works only in the long cylinder limit, and consequently the associated calculations are exponentially costly\cite{Jiang2012TEE}.

\begin{figure}
\centerline{
    \includegraphics[height=4.4in,width=3.4in] {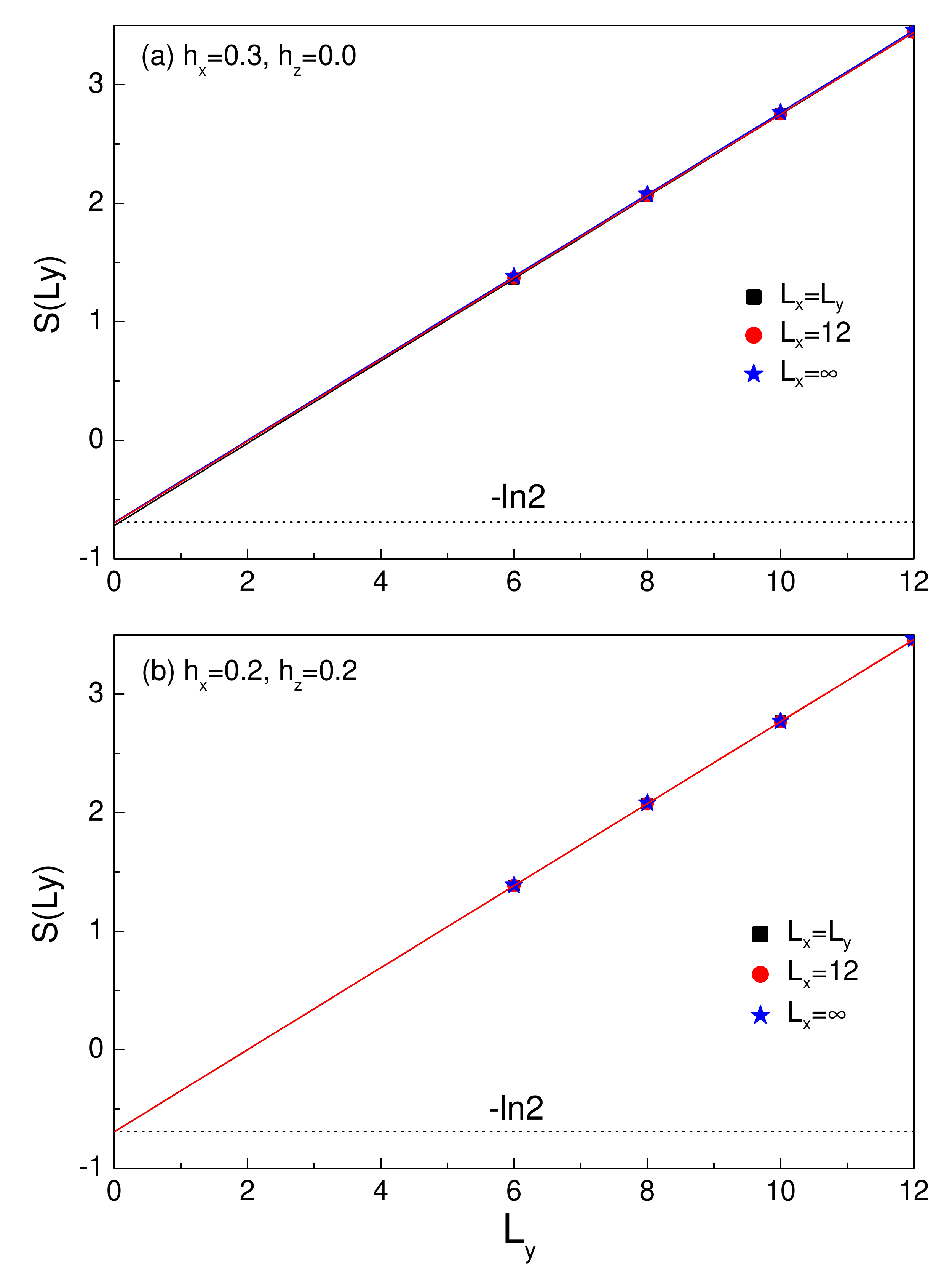}
    }
\caption{(Color online) The von Neumann entanglement entropy
$S(L_y)$ for the Toric-Code model on a cylinder with $L_y=6-12$ at $L_x=L_y$
(black squares), $L_x=12$ (red circles), and $L_x=\infty$ (blue stars), for (a) $h_x=0.3$, $h_z=0.0$, and (b) $h_x=h_z=0.2$. By fitting $S(L_y)=aL_y-\gamma$, we get $\gamma=\ln(2)$ to high accuracy for all cases, i.e., $\sim 3\%$ ($L_x=L_y$) and $\sim 1\%$ ($L_x=12$) for (a), and $\sim 0.2\%$ ($L_x=L_y$) and $\sim 0.02\%$ ($L_x=12$) for (b) The other parameters for these runs are $\varepsilon_r=10^{-6}$, $n_q=4$, $\varepsilon_q=5\times 10^{-3}$ for $L_y\leq 8$, and $\varepsilon_q=10^{-3}$ for $L_y\geq 10$.}
\label{Fig:TCMMESEntropy}
\end{figure}

To avoid this exponential cost, it is clearly desirable to obtain a MES for finite systems.  This is not possible in the standard DMRG method (Fig.\ref{Fig:DMRGPorcedure}(a)), as is clear from several examples.  One seeks, following Ref.\cite{Jiang2012TEE}, a value for the minimum number of states $m_0$ which is large enough that convergence of local entanglement is achieved but small enough that the non-local absolute ground state is not yet found, i.e. the entanglement entropy is that of the MES.  We see in Fig.\ref{Fig:TCMHx03Hz00MS}(a) (black squares), that for a system of size $N=L_x \times L_y=12\times 6$, the standard DMRG finds the proper entanglement entropy of the MES only for a very narrow range of the number of states $m_0$.  The range of convergence to the appropriate MES entropy becomes larger with increasing system width, as shown for $N=10\times 10$ in Fig.\ref{Fig:TCMHx03Hz00MS}(b), and in this case we could choose by this observation an appropriate value of $m_0$ to obtain an MES. However, if we decrease the system size, no entropy plateau at all appears, as seen for $N=6\times 6$ in Fig.\ref{FigS:TCMN6x6MES}(b).

The improved procedure, by contrast, gives the MES easily and systematically for finite systems.   In Fig.~\ref{Fig:TCMHx03Hz00MS}, we fix $m_1/m_0\leq 1$ and vary $m_0$, calculating the entanglement entropy $S$.  With this protocal, a large plateau of $S$ versus $m_0$ is observed at the correct value for the MES.  The plateau occurs for $m_0>m_{0c}$, with $m_{0c}$ decreasing somewhat with increasing $m_1/m_0$.  The same behavior is observed for both $N=12\times 6$ and $10\times 10$.  Remarkably, the entanglement entropy measured in the MES determined this way is almost {\em independent} of the cylinder length $L_x$, and thus equal to the infinite cylinder value.  This is shown in Fig.~\ref{Fig:TCMMESEntropy}.  Hence by obtaining the MES in this fashion, we can avoid simulating long cylinders at all and obtain accurate values for the topological entanglement entropy from simulations with, for example, $L_x=L_y$.

\begin{figure}
\centerline{
    \includegraphics[height=4.2in,width=3.4in] {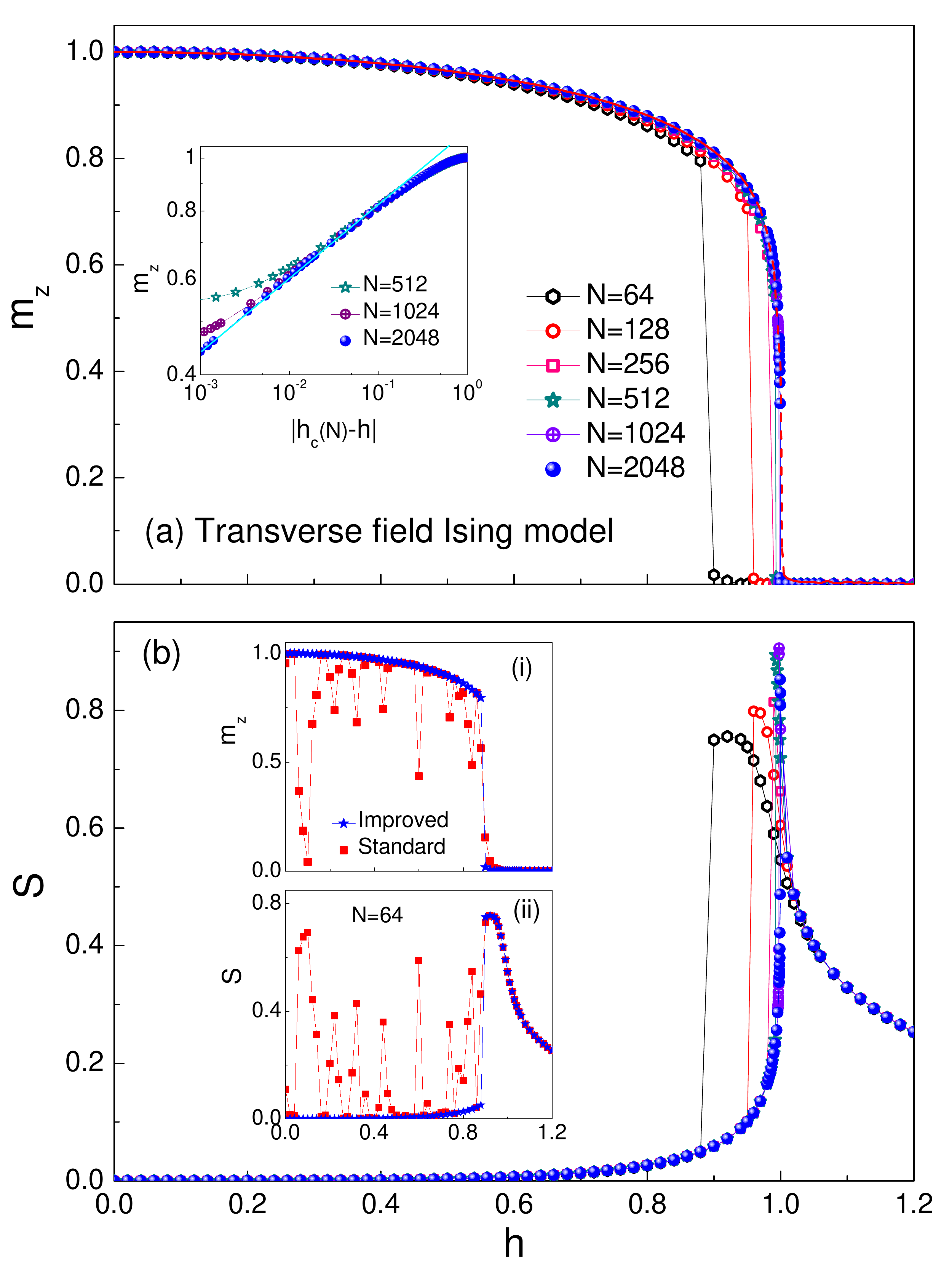}
    }
\caption{(Color online) (a) Magnetization $m_z=|\langle\sigma^z\rangle|$ and (b) von Neumann entanglement entropy $S$ versus the transverse magnetic field $h$ for different system sizes $N$ in the transverse field Ising model [Eq.(\ref{Eq:TransverseIsingModel})] on an open chain, obtained using the improved procedure with $m_1=m_0$. Inset in (a): log plot for $m_z=A|h_c(N)-h|^\beta$, where $h_c(N)$ is the critical field for system $N$ (see text). After fitting, we obtain $\beta=0.128(4)$ for $N=2048$. Insets in (b): (i) magnetization $m_z$ and (ii) entanglement entropy $S$ for $N=64$, obtained using the standard procedure (red squares) and improved procedure (blue stars). Here other parameters are $m_0=64$,  $n_q=4$, $\varepsilon_q=5\times 10^{-1}$ and $\varepsilon_r=10^{-8}$.} \label{Fig:IsingChain}
\end{figure}

\section{Transverse-field Ising model}
\label{Sec:IsingChain}

In the previous section, we saw that the improved DMRG procedure avoids non-local entanglement and obtains the MES reliably for the Toric-Code model in its topologically ordered phase.   Now we turn to the ``opposite'' type of non-local entanglement, which arises due to the formation of a Schr\"odinger cat state in a broken symmetry phase.  We consider in this section the canonical one-dimensional transverse field Ising model,
\begin{eqnarray}
H=-\sum_{i}\sigma^z_i\sigma^z_{i+1} -
h\sum_i\sigma^x_i,\label{Eq:TransverseIsingModel}
\end{eqnarray}
where $\sigma^x_i$ and $\sigma^z_i$ are Pauli matrices on site $i$. This model is known to have a topological trivial ground state in all magnetic fields, and a second order phase transition at the critical field $h_c=1.0$\cite{SachdevPhysToday}. Below the critical point, i.e., $h<h_c$, the ground state is a ferromagnet with all spins parallel along $z$-axis, which spontaneously breaks the $Z_2$ symmetry $\sigma_i^z \rightarrow - \sigma_i^z$, $\sigma_i^y \rightarrow - \sigma_i^y$.

As for the TCM, the standard procedure converges to an apparently random superposition of the two-fold quasi-degenerate states (a generic such superposition is a ``partial Schr\"odinger cat''). Unlike for the case of topological order, the different degenerate states are here locally distinguishable, which leads to more dramatic signatures of this arbitrariness.  Specifically, not only does entanglement entropy sense this superposition, so does the magnetization.  As shown in the insets of Fig.~\ref{Fig:IsingChain}, the magnetization and entanglement entropy measured by the standard DMRG indeed fluctuate strongly. 

By contrast, the improved procedure nicely produces a MES with non-fluctuating magnetization $m_z=|\langle \sigma^z\rangle|$ and entanglement entropy.  Note that in fact the sign of the magnetization fluctuates (so we measure its absolute value) as the improved simulation has no bias toward either symmetry broken state, only toward low entropy states.  Still, a direct measure of the order parameter is possible: we measure this ``property of the thermodynamic limit'' in a finite system!   This is a distinct advantages over other approaches to compute the magnetization, such as including a pinning field (which may bias the results) or extrapolating from spin-spin correlation functions (which increases error since short-range contributions must be removed and a square root taken).    

To demonstrate the accuracy of the method, we obtain the critical exponent for the magnetization. The phase transition point for the finite system, $h_c(N)$, is determined from the simulation as the field value corresponding to the mid-point of the rapid drop in magnetization (which can be seen in Fig.\ref{Fig:IsingChain}).  The critical field converges toward $h_c(\infty)=1$, as expected. Then by fitting the slope on a log-log plot of $m_z$ versus $|h_c(N) - h|$, we obtain $\beta=0.128(4)$ for $N=2048$, close to the theoretical value $\beta=\frac{1}{8}$.

\begin{figure}
\centerline{
    \includegraphics[height=4.2in,width=3.4in] {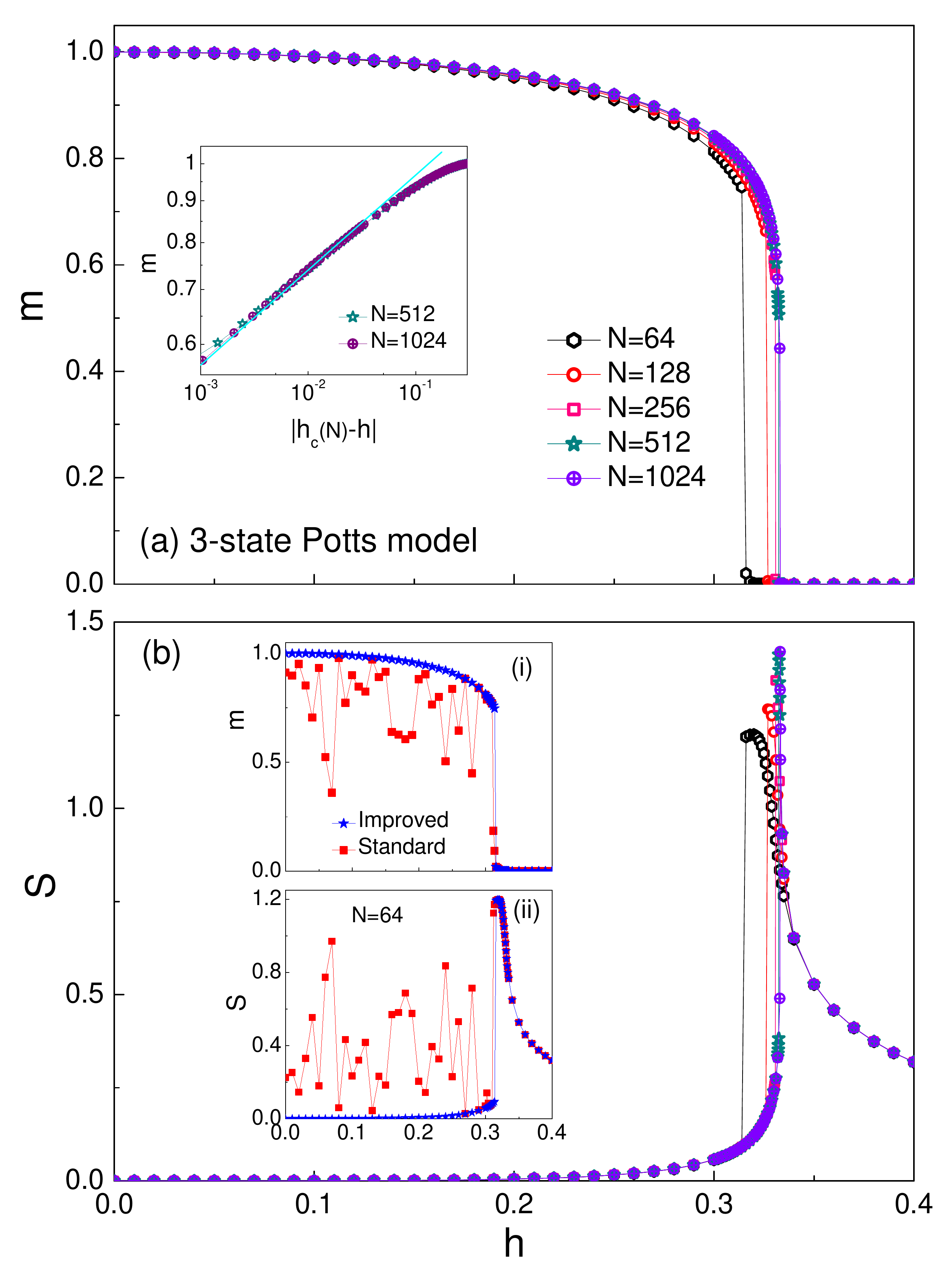}
  } \caption{(Color online) (a) Magnetization $m=|\langle\Psi\rangle|$ and (b) von Neumann entanglement entropy $S$ versus the magnetic field $h$ for different system sizes $N$ for the $3-$state Potts model [Eq.(\ref{Eq:PottsModel})] on an open chain, obtained using the improved procedure with $m_1=m_0$. Inset in (a): log plot of $m_z=A|h_c(N)-h|^\beta$, where $h_c(N)$ is the critical field for system $N$. After fitting, we find $\beta=0.110(1)$ for $N=1024$. Insets in (b): (i) magnetization $m$ and (ii) entanglement entropy $S$ for $N=64$, obtained using the standard procedure (red squares) and improved procedure (blue stars). Here other parameters are $m_0=64$, $n_q=4$, $\varepsilon_q=5\times 10^{-1}$ and $\varepsilon_r=10^{-8}$.} \label{Fig:PottsModel} \end{figure}

\section{Three-state Potts model}
\label{Sec:PottsModel}

Since the Ising model is especially simple and may exhibit some non-generic features, we consider another example of a system with spontaneous symmetry breaking.  This is the $q$-state Potts model, which consists of a lattice of spins, which can take $q$ different values, and the Hamiltonian in one dimension is given by
\begin{eqnarray}
H=-\sum_{i}\sum_{n=1}^q\sigma^n_i \sigma^n_{i+1} - h\sum_{i}Q_i.\label{Eq:PottsModel}
\end{eqnarray}
For $q=2$, the Potts model is equivalent to the Ising model in Eq.(\ref{Eq:TransverseIsingModel}). In this section, we will focus on  the the $q=3$ case, i.e., the 3-state Potts model, which has 3-fold ground state degeneracy. The matrix elements of the operators are given by 
\begin{eqnarray}
  \label{eq:1}
  \left(\sigma^n\right)_{ab} & = & \delta_{an}\delta_{bn}, \\
  \left( Q\right)_{ab} & = & 1 - \delta_{ab},
\end{eqnarray}
with $a,b \in \{1,2,3\}$.  Eq.~(\ref{Eq:PottsModel}) is equivalent to the integrable 3-state Potts chain in for example Ref.\cite{Gehlen1986Q3PottsModel} by a unitary transformation, from which one may deduce the critical point lies at $h_c=1/3$ in the infinite system.

We carried out the same measurements for the Potts model as we did for the Ising chain.  Here the order parameter is defined by $m=|\langle\Psi_i\rangle|$, where the magnetization operator on a single site is
\begin{equation}
  \label{eq:2}
  \left( \Psi \right)_{ab} = e^{2\pi i (a-1)/3}\delta_{ab}.
\end{equation}
The results for the magnetization and entanglement entropy are shown in Fig.~\ref{Fig:PottsModel}.   We observe behavior consistent with the thermodynamic limit and the exact location of the critical point at $h_c=1/3$. From the numerical determination of the phase transition point for finite $N$, the critical exponent $\beta=0.110(1)$ is found by fitting $M=A|h_c(N)-h|^\beta$ for $N=1024$, close to the theoretical value $\beta=\frac{1}{9}$.

\begin{figure}
\centerline{
    \includegraphics[height=2.4in,width=3.4in] {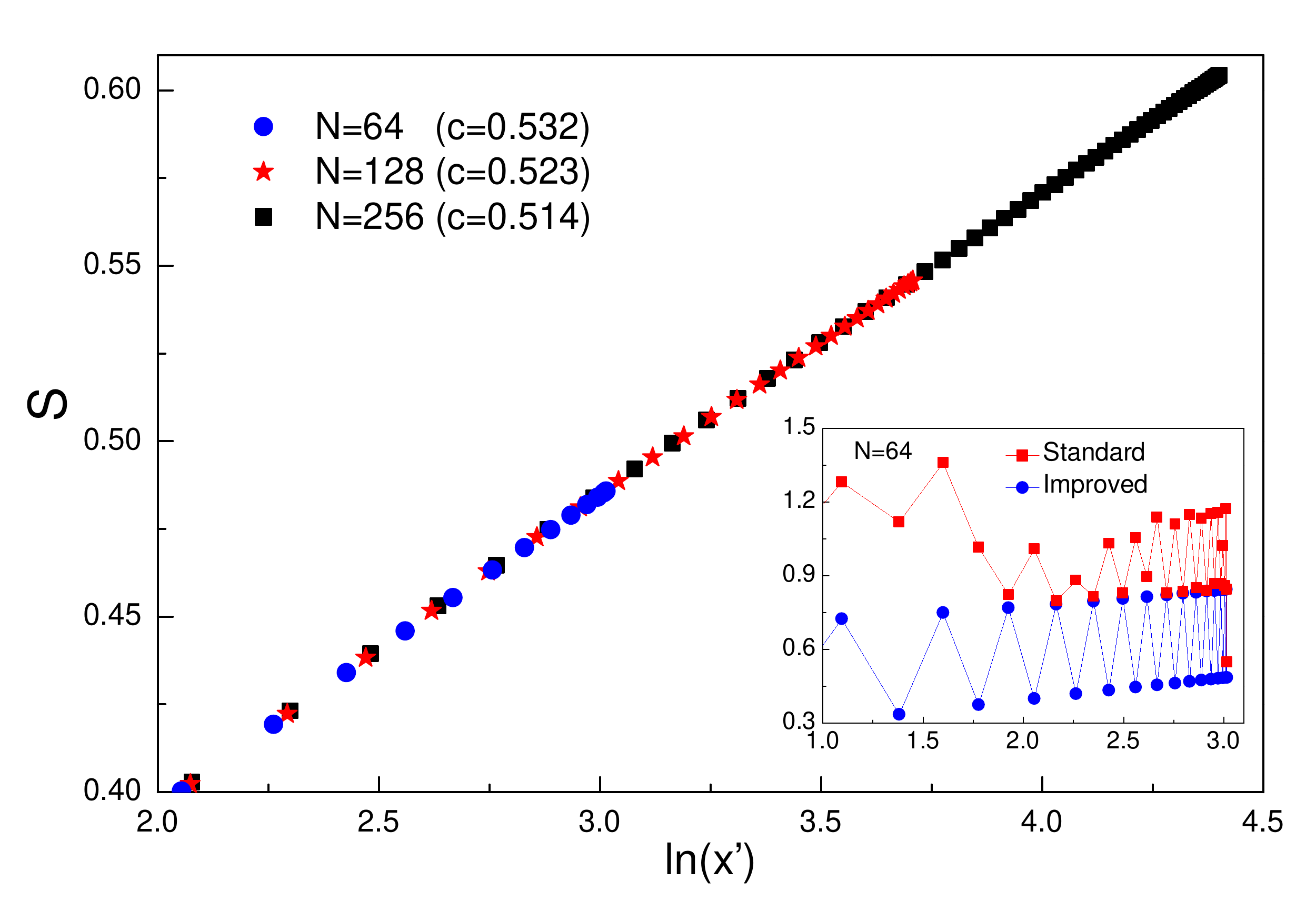}
    }
\caption{(Color online) The lower branch of the von Neumann
entanglement entropy $S$ of the open Kitaev chain in Eq.(\ref{Eq:MajoranaChain}), obtained using the improved procedure with $m_1=m_0$. Here $c$ is the central charge fitted by the formula $S=\frac{c}{6}\rm{ln}(x')+const$, where $x'=\frac{N}{\pi}\sin(\frac{\pi x}{N})$, and $x$ is the length of the subsystem. Inset: The full entanglement entropy $S$ obtained using the standard procedure (red squares) and improved procedure (blue circles) for $N=64$. The other  parameters used in these calculations are $m_0=64$,  $n_q=4$, $\varepsilon_q=5\times 10^{-1}$ and $\varepsilon_r=10^{-8}$.} \label{Fig:MajoranaChain}
\end{figure}

\section{Majorana Fermion Chain}
\label{Sec:MajoranaChain}

As a final illustration of the method, we consider the rather peculiar example of the Kitaev alternating chain, whose Hamiltonian is given by
\begin{eqnarray}
H=-\sum_{i}^{N/2}\left(\sigma^x_{2i-1}\sigma^x_{2i}+\sigma^y_{2i}\sigma^y_{2i+1}\right).
\label{Eq:MajoranaChain}
\end{eqnarray}
Here $\sigma^x_i$ and $\sigma^y_i$ are Pauli matrices on site $i$, and $N$ is the number of sites.   One can easily show by a Jordan-Wigner transformation that this model is equivalent to a critical massless Majorana fermion chain {\em and} a decoupled set of $N$ localized Majorana fermions which do not enter the Hamiltonian at all\cite{Feng2007KitaevChain}.\ From the point of view of conformal field theory, the massless Majorana fermion is a central charge $c=1/2$ theory, but in addition the localized Majorana fermions induce a massive $2^{N/2-1}$-fold degeneracy for an open chain, which is decoupled from the critical modes.  We ask here whether it is possible to measure the $c=1/2$ conformal central charge numerically.

For the standard DMRG, this is an insurmountable challenge, as it is very easy to generate ``spurious'' entanglement (i.e. not associated with the conformal field theory) from the massive ground state degeneracy.  This renders the corresponding entropy calculated in the standard way highly and randomly fluctuating (see inset of  Fig.~\ref{Fig:MajoranaChain}).  On the contrary, the improved procedure picks one of the MES ground states, despite the huge ground state degeneracy.  All the remaining entanglement is due to the conformal field theory, and from it we determine correctly $c=1/2$ (with approximately 2 percent error) by standard methods of analysis (main panel of Fig.~\ref{Fig:MajoranaChain}) for extracting $c$ from the entanglement entropy.

\section{Summary and Conclusion}%

In this paper, we proposed an improved DMRG procedure to remove unnecessary non-local entanglement and obtain the minimally entangled states which are naturally selected by decoherence of large physical systems. In such a MES, intrinsic properties of the thermodynamic limit are directly obtained already for finite systems.  We showed by example that the method is successful for various gapped systems such as the two-dimensional toric code model and two models of spontaneously broken discrete symmetries in one dimension, as well as for the unusual Kitaev alternating chain, for which an extensive degeneracy and conformal criticality coexist.  It is our expectation that the method should be successful for generic gapped Hamiltonians in one dimensional or quasi-one-dimensional geometries.  It would be interesting in the future to try its application to more complicated gapped systems with complex translational symmetry breaking, e.g. valence bond solids with large unit cells.   A generalization to systems with continuous symmetry breaking is also desirable, but likely requires further theoretical development.

\emph{\textbf{Acknowledgement}} HCJ thanks Steven Kivelson, Xiao-Liang Qi, Srinivas Raghu, Meng Cheng, Zhenghan Wang and Senthil Todadri for helpful discussions. We acknowledge computing support from the Center for Scientific Computing at the CNSI and MRL: an NSF MRSEC (DMR-1121053) and NSF CNS-0960316. This research was supported by the NSF through grant DMR-12-06809.


\appendix






\section{DMRG Basics}%
\label{appendixDMRG}%

\renewcommand{\thefigure}{A\arabic{figure}}
\setcounter{figure}{0}
\renewcommand{\theequation}{A\arabic{equation}}
\setcounter{equation}{0}

In this section, we will review relevant aspects of the standard DMRG algorithm, including DMRG truncation, infinite-system DMRG algorithm, and finite-system DMRG algorithm, which can be found in Ref.\cite{White1992DMRG,White1993DMRG,DMRG_RMP,Schollwock2011DMRG}

\begin{figure} \centerline{\includegraphics[height=1.6in,width=2.2in]{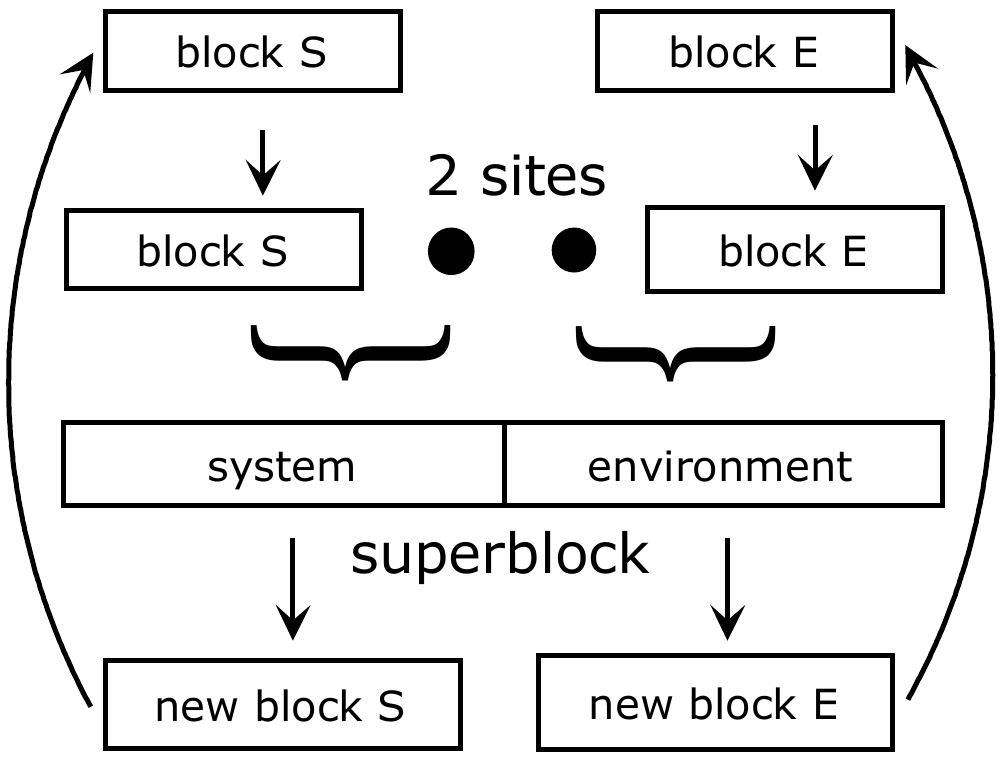}}%
  \caption{Infinite-system DMRG algorithm, i.e., the warm-up process.  The diagram shows the fundamental DMRG construction of a superblock from two blocks and two single central sites. See Ref. \cite{White1992DMRG,White1993DMRG,DMRG_RMP,Schollwock2011DMRG} for more details.}
\label{FigS:DMRGWarmup}
\end{figure}

In the standard DMRG approach, to find the states to be kept and obtain any reasonable approximation to the reduced density matrix, one needs to diagonalize the Hamiltonian of a larger system, called the superblock (as shown in Fig.\ref{FigS:DMRGWarmup}), which includes a
\emph{system} block $\rm S^\prime$, i.e., the block S and the left central site, and an \emph{environment} block $\rm E^\prime$, i.e., the block E and the right central site. The ground state for the superblock is obtained using the Lanczos algorithm of matrix diagonalization, namely $|\psi\rangle=\sum_{ij}\psi_{ij}|i\rangle|j\rangle$. Here $|i\rangle$, ($i=1,\cdots,M$) and $|j\rangle$, ($j=1,\cdots,J$) are the complete sets of states of the \emph{system} $\rm S^\prime$ and
\emph{environment} $\rm E^\prime$, respectively. Then we can form the reduced density matrix $\rho={\rm Tr}_{E}|\psi\rangle\langle\psi|$ for the \emph{system} and determine its eigenstates $|\lambda_\alpha\rangle$ ordered by descending eigenvalues (weight) $\lambda_\alpha$, with $\sum_\alpha \lambda_\alpha=1$ and $\lambda_\alpha\geq0$. The optimal states are the eigenstates $|\lambda_\alpha\rangle$ of $\rho$ with the largest eigenvalues.  The truncation error can then be defined as
\begin{eqnarray}
\varepsilon\equiv \sum_{\alpha=m+1}^M\lambda_\alpha,
\label{EqS:DMRGTruncError}
\end{eqnarray}
which gives the deviation of $P_m=\sum_{\alpha=1}^m\lambda_\alpha$ from unity, and measures the accuracy of the truncation to $m$ states. A similar procedure applies to the environment.

The superblock configuration mentioned above can be used in two different ways, for the infinite-system DMRG algorithm and the finite-system DMRG algorithm. Fig.~\ref{FigS:DMRGWarmup} shows the infinite-system DMRG algorithm, i.e., the \emph{warm-up} process.  This is generally used to build up the system to a given size, and serves as the initial step of the finite-system algorithm. In the warm-up process, a tentative new system block $\rm S^\prime$ (with Hamiltonian $H^\prime$) is formed from the block S and one added site (central site), and a new tentative
environment block $\rm E^\prime$ is built from block E in the same way.   Then one builds the superblock from $\rm S^\prime$ and $\rm E^\prime$, and diagonalizes it to obtain the ground state
wavefunction $|\psi\rangle$. Subsequently, the reduced density matrix $\rho={\rm Tr}_{E'}|\psi\rangle\langle\psi|$ and its eigenstates $|\lambda_\alpha\rangle$ are determined, which defines the projection operator $T= \sum_{\alpha=1}^m |\lambda_\alpha\rangle \langle\lambda_\alpha|$. One then carries out the reduced basis transformation $H^{tr}=T^{+}H^\prime T$ onto the new $m$-state basis to form a new block S and new block E. Finally, the original blocks S and E are replaced by the new ones, and the steps are repeated until the desired system size is reached.

\begin{figure}
\centerline{\includegraphics[height=2.4in,width=3.4in]{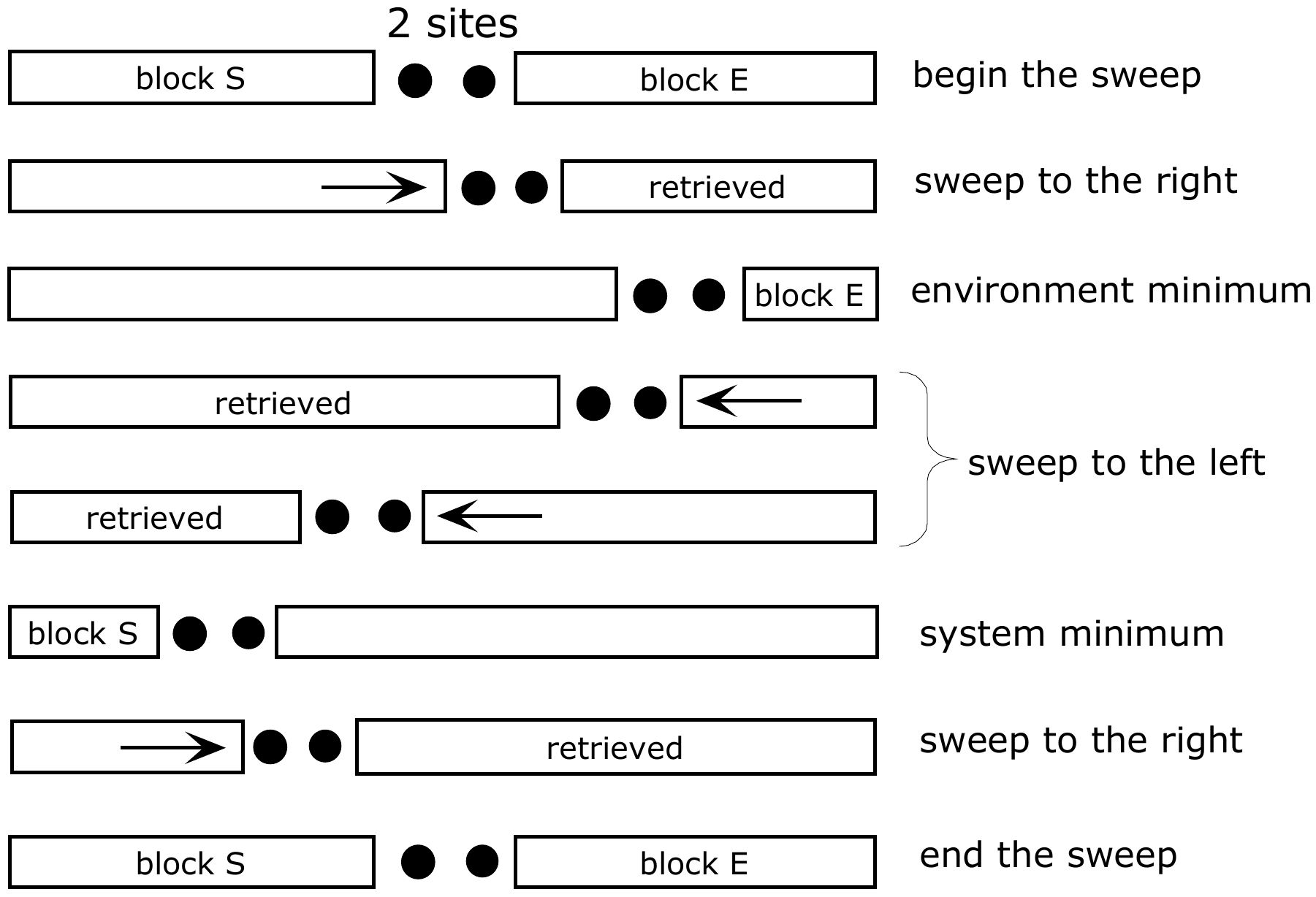}}%
\caption{Sweep process, i.e., the finite-system DMRG algorithm.  The diagram shows the process of progressive block growth and shrinkage. Here, ``begin the sweep" can either be the "end of infinite-system DMRG" or the ``end of one sweep". See Ref.\cite{White1992DMRG,White1993DMRG,DMRG_RMP,Schollwock2011DMRG} for more details.} \label{FigS:DMRGSweep}
\end{figure}

The infinite system DMRG has not been proven satisfactorily accurate in many cases, for example for the computation of correlation functions.  The finite-system DMRG algorithm was invented and designed to eliminate this concern, and to calculate accurately the properties of a finite system of size $N$. The idea of the finite-system algorithm is to stop the infinite-system algorithm at some preselected superblock length $N$ which is kept fixed. In subsequent DMRG steps, as shown in Fig.\ref{FigS:DMRGSweep}, one applies the steps of infinite-system DMRG, but instead of simultaneous growth of both blocks, growth of one block is accompanied by shrinkage of the other block. Reduced basis transformations are carried out only for the growing block.   As the system block grows at the expense of the environment block, environment blocks of all sizes and their operators must have been stored previously (at the infinite-system DMRG stage or during previous iterations of finite-system DMRG). When the environment block reaches some minimum size, it can be treated exactly, and the growth direction is reversed. The environment block now grows at the expense of the system block. All basis states are chosen while the system and environment are embedded in the final system and with the knowledge of the full Hamiltonian. A complete shrinkage and growth sequence for both blocks is called a \emph{sweep}, as shown in Fig.\ref{FigS:DMRGSweep}. Finally, converged results can be obtained by performing enough sweeps with a sufficient number of states kept.

\section{Toric-Code Model in Magnetic Field}%
\label{appendixTCM}%
\renewcommand{\thefigure}{B\arabic{figure}}
\setcounter{figure}{0}
\renewcommand{\theequation}{B\arabic{equation}}
\setcounter{equation}{0}

\begin{figure}
\centerline{\includegraphics[height=4.2in,width=3.2in]{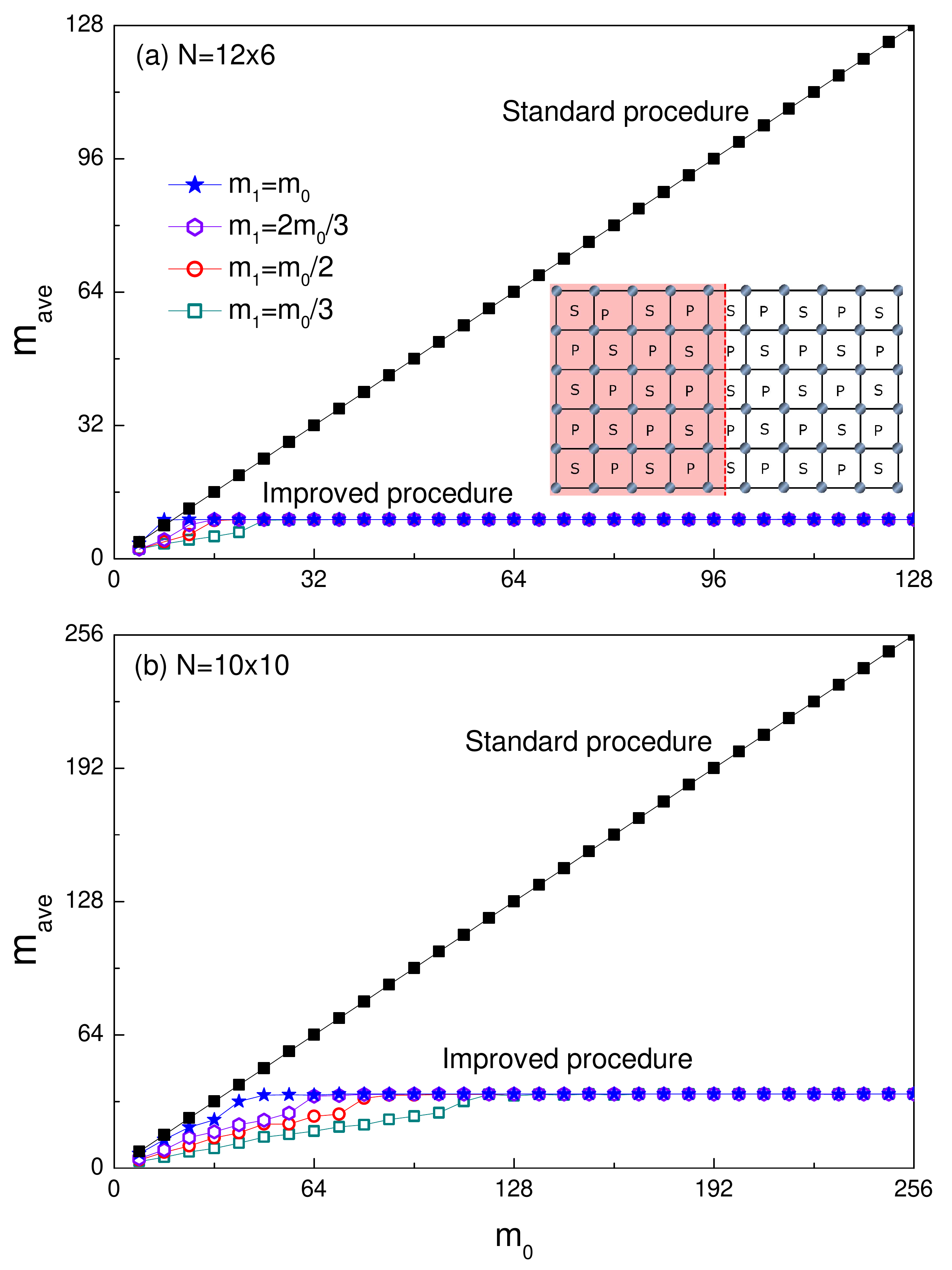}}%
\caption{(Color online) The average number of states $m_{ave}$ that is needed to obtain a MES versus $m_0$ at different $m_1$ for the TCM (see Fig.\ref{Fig:TCMHx03Hz00MS}) on a cylinder with $h_x=0.3$ and $h_z=0.0$, for (a) $N=12\times 6$ and (b) $N=10\times 10$. Black squares denote the results obtained using the standard procedure, and the rest are obtained using the improved procedure. Here other parameters are $\varepsilon_r=10^{-6}$, $n_q=4$, $\varepsilon_q=5\times 10^{-3}$ for $N=12\times 6$, and $\varepsilon_q=10^{-3}$ for $N=10\times 10$. Inset in (a): Square lattice with $L_x=10$ and $Ly=6$. Here $S$ represents the star operator $A_s$, while $P$ represents the plaquette operator $B_p$.} \label{FigS:TCMM0MaveMES}
\end{figure}

The Toric-Code model\cite{Kitaev2003} with an applied magnetic field is given by
\begin{eqnarray}
&& H=-J_s\sum_s A_s - J_p\sum_p B_p - h_x\sum_i\sigma^x_i -
h_z\sum_i\sigma^z_i,\nonumber \\
&&
\label{EqS:ToricCodeModel}
\end{eqnarray}
where $\sigma^x_i$ and $\sigma^z_i$ are Pauli matrices, $A_s=\Pi_{i\in   s}\sigma^x_i$ and $B_p=\Pi_{i\in p}\sigma^z_i$.  Subscripts $s$ and $p$ refer respectively to vertices and plaquettes on the square lattice, whereas $i$ runs over all bonds where spin degrees of freedom are located. Without magnetic field, i.e., $h_x=h_z=0$, the pure TCM can be solved exactly\cite{Kitaev2003}.  It exhibits a 4-fold ground state degeneracy on the torus, and has $Z_2$ topological order with total
quantum dimension $D=2$. All elementary excitations are gapped and characterized by eigenvalues $A_s=-1$ (a $Z_2$ charge on site $s$) and $B_p=-1$ (a $Z_2$ vortex on plaquette $p$). After turning on the magnetic fields, the model cannot be solved exactly anymore. However, previous studies\cite{Trebst2007,Vidal2008,Tupitsyn2008} show that the $Z_2$ topological phase remains quite stable and robust until the magnetic fields are large enough to induce a phase transition from the topological phase to the trivial one.  Specifically, such a phase transition takes place at the critical magnetic field $h_x^c\approx 0.328$ when $h_z=0$, while $h_x^c\approx 0.34$ along the symmetric line with $h_z=h_x$.

\begin{figure}
\centerline{\includegraphics[height=4.0in,width=3.2in]{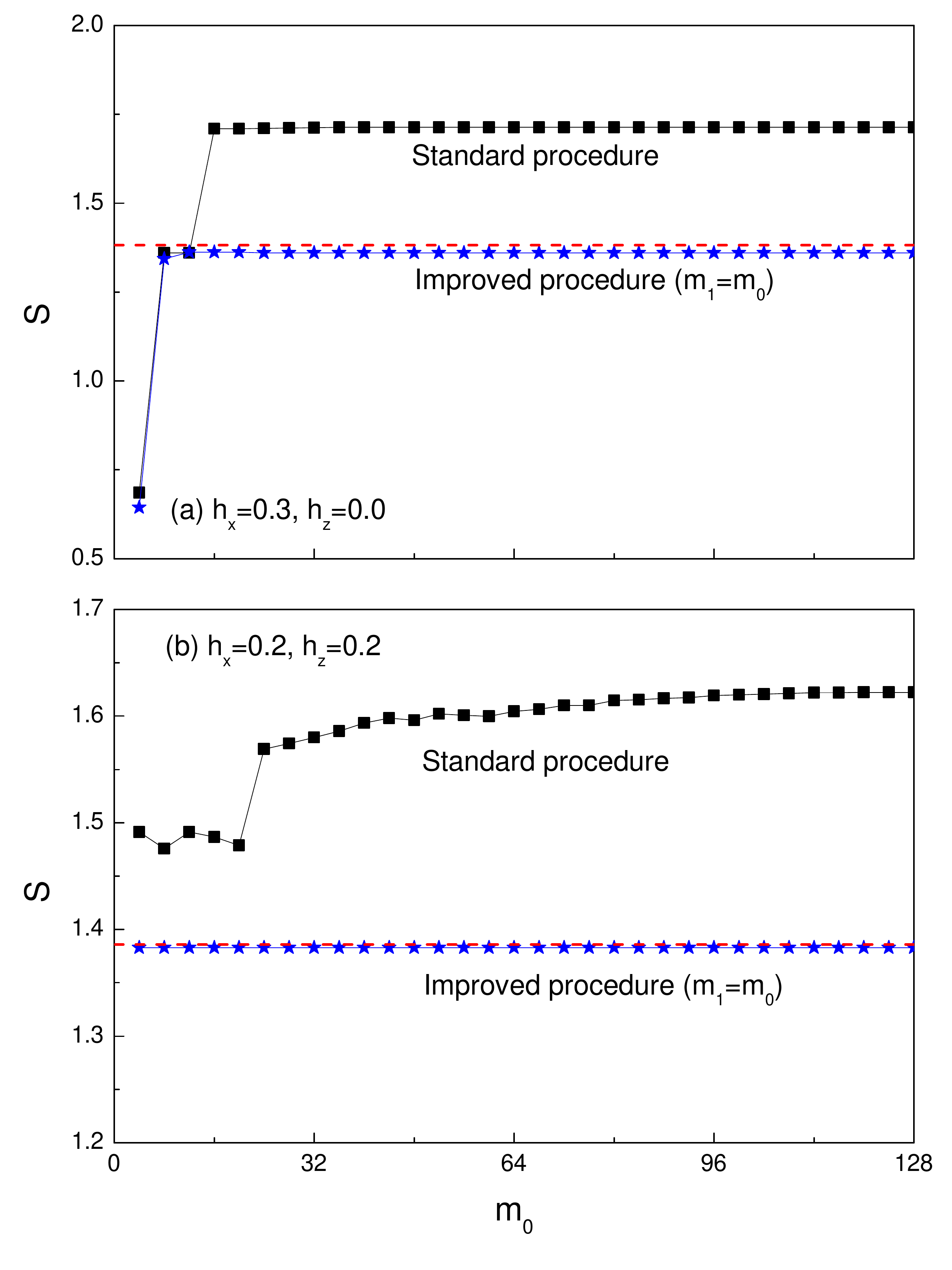}}%
\caption{(Color online) The von Neumann entanglement entropy $S$ versus
the number of states $m_0$ for the TCM on a cylinder with $N=6\times 6$ at (a) $h_x=0.3$ and $h_z=0.0$, and (b) $h_x=h_z=0.2$. Red dashed lines denote the results in the
long-cylinder limit $L_x=\infty$, black squares denote the results obtained using the standard procedure, and the blue stars are obtained using the improved procedure with $m_1=m_0$. Here other parameters are $\varepsilon_r=10^{-6}$, $n_q=4$ and $\varepsilon_q=5\times 10^{-3}$.}
\label{FigS:TCMN6x6MES}
\end{figure}

For the DMRG simulation, we consider an equivalent square lattice, where the spin operators $\sigma^x$ and $\sigma^z$ sit on the sites instead of the bonds. Therefore, the star operator $A_s$ and the plaquette operator $B_p$ of the original lattice now sit on alternating plaquettes in the equivalent square lattice, as shown in the inset of Fig.~\ref{FigS:TCMM0MaveMES}(b), labeled as $S$ and $P$, respectively. In the main text, we have systematically computed the von Neumann entanglement entropy for the TCM for two different system sizes, i.e., $N=12\times 6$ and $N=10\times 10$, using both standard procedure and improved procedure. We find that the standard procedure generally produces a high entropy state (HES), instead of a MES. The region with the MES and correct entanglement entropy, as a function of the number of states $m_0$, is usually very tiny or even vanishing, if the system is not big enough. On the contrary, the improved procedure gives the MES in a much larger region, even for small systems. Examples can be found in Fig.\ref{Fig:TCMHx03Hz00MS} in the main text, and Fig.\ref{FigS:TCMN6x6MES} in the Supplemental Information.

\begin{figure}
\centerline{\includegraphics[height=4.0in,width=3.2in]{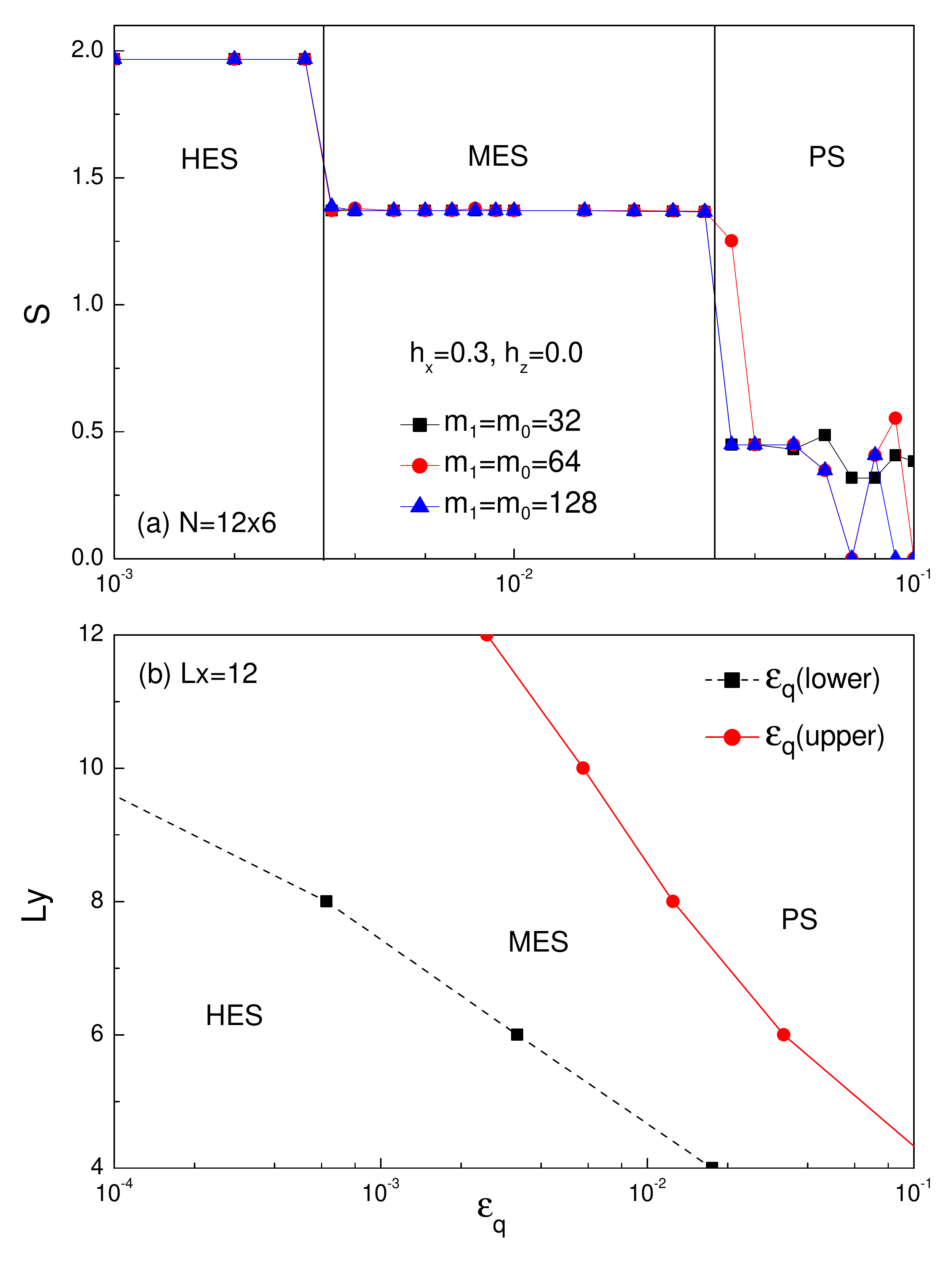}}%
\caption{(Color online) (a) The von Neumann entanglement entropy $S$ versus the truncation error $\varepsilon_q$ for the TCM on a cylinder with $N=12\times 6$ and $h_x=0.3$ ,$h_z=0.0$, obtained using the improved procedure with different $m_1=m_0$. (b) The ``phase diagram'' for states found using the improved procedure as a function of the truncation error in the quench process, $\varepsilon_q$, and the system width $L_y$ for the TCM with $L_x=12$. HES, MES and PS denote high entropy state, minimally entangled state and product-like state, respectively. Here other parameters are $\varepsilon_r=10^{-6}$ and $n_q=4$.} \label{FigS:TCMNErrQMES}
\end{figure}

Here, we discuss in more detail how the improved procedure works for the TCM. In particular, we compute the actual number of states and the average value $m_{ave}$ that are needed to obtain the MES in the \emph{recovery} process. Examples are given in Fig.\ref{FigS:TCMM0MaveMES} for both $N=12\times 6$ and $N=10\times 10$, the same systems shown in Fig.\ref{Fig:TCMHx03Hz00MS}. 
From the figure we can see that $m_{ave}$ does not depend on $m_1$ and $m_0$, once the system enters into the MES, although the critical value $m_{0c}$ that is needed for the MES depends on $m_1/m_0$. Generally, $m_{ave}$ is smaller or much smaller than the $m_0$ which is used in the standard procedure, and can be determined automatically by the truncation error $\varepsilon_r$ in the improved procedure. For example, for $\varepsilon_r=10^{-6}$, $m_{ave}\approx 9.4$ for $N=12\times 6$, and $m_{ave}\approx 35.6$ for $N=10\times 10$ for $h_x=0.3$ and $h_z=0.0$.

For smaller systems, it may become difficult or impossible to obtain an MES using the conventional algorithm.  This is the case for the system of size $N=6\times 6$, for which data is shown in Fig.\ref{Fig:TCMMESEntropy} and Fig.\ref{FigS:TCMN6x6MES}.  For the parameters $h_x=h_z=0.2$, no MES is found by the standard algorithm. On the contrary, the improved procedure obtains it for all cases studied.

Finally, we discuss the choice of truncation error $\varepsilon_q$ in the \emph{quench} process. As was mentioned in the main text, the appropriate value for $\varepsilon_q$ is slightly dependent on system width $L_y$, and hence on the entanglement entropy. We generally take $\varepsilon_q$ larger for thinner systems with smaller entanglement entropy, and smaller for wider systems with larger entanglement entropy.  Examples of the performance of the simulation with respect to $\varepsilon_q$ in the TCM with $L_x=12$ are shown in Fig.\ref{FigS:TCMNErrQMES}.  For a given width $L_y=6$, i.e., $N=12\times 6$, we computed the entanglement entropy as a function of $\varepsilon_q$ for given $m_0$ and $m_1$, as shown in Fig.\ref{FigS:TCMNErrQMES}(a). When $\varepsilon_q$ is smaller than a lower bound, i.e., $\varepsilon_q<\varepsilon_{q1}$, and $m_0$ and $m_1$ are large, we obtain a HES, such as the equal-weight linear superposition of two MESs. On the contrary, when $\varepsilon_q$ is larger than some upper bound, i.e., $\varepsilon_q>\varepsilon_{q2}$, the system loses its entanglement with the environment and becomes a product-like state (PS). Note that both $\varepsilon_{q1}$ and $\varepsilon_{q2}$ depend slightly on $m_0$ and $m_1$. Interestingly, there is a finite region between HES and PS, in which the system converges to a MES. The ``phase diagram'' showing the regions of convergence to a HES, MES, and PS in the space of truncation error $\varepsilon_q$ and system width $L_y$ is plotted in Fig.\ref{FigS:TCMNErrQMES}(b) for the TCM with $L_x=12$.  We see a wide region in which a MES is found, which is the desired parameter space for a successful simulation.


\begin{thebibliography}{23}
\expandafter\ifx\csname natexlab\endcsname\relax\def\natexlab#1{#1}\fi
\expandafter\ifx\csname bibnamefont\endcsname\relax
  \def\bibnamefont#1{#1}\fi
\expandafter\ifx\csname bibfnamefont\endcsname\relax
  \def\bibfnamefont#1{#1}\fi
\expandafter\ifx\csname citenamefont\endcsname\relax
  \def\citenamefont#1{#1}\fi
\expandafter\ifx\csname url\endcsname\relax
  \def\url#1{\texttt{#1}}\fi
\expandafter\ifx\csname urlprefix\endcsname\relax\def\urlprefix{URL }\fi
\providecommand{\bibinfo}[2]{#2}
\providecommand{\eprint}[2][]{\url{#2}}

\bibitem[{\citenamefont{White}(1992)}]{White1992DMRG}
\bibinfo{author}{\bibfnamefont{S.~R.} \bibnamefont{White}},
  \bibinfo{journal}{Phys. Rev. Lett.} \textbf{\bibinfo{volume}{69}},
  \bibinfo{pages}{2863} (\bibinfo{year}{1992}).

\bibitem[{\citenamefont{White}(1993)}]{White1993DMRG}
\bibinfo{author}{\bibfnamefont{S.~R.} \bibnamefont{White}},
  \bibinfo{journal}{Phys. Rev. B} \textbf{\bibinfo{volume}{48}},
  \bibinfo{pages}{10345} (\bibinfo{year}{1993}).

\bibitem[{\citenamefont{Schollw\"ock}(2005)}]{DMRG_RMP}
\bibinfo{author}{\bibfnamefont{U.}~\bibnamefont{Schollw\"ock}},
  \bibinfo{journal}{Rev. Mod. Phys.} \textbf{\bibinfo{volume}{77}},
  \bibinfo{pages}{259} (\bibinfo{year}{2005}).

\bibitem[{\citenamefont{Schollw枚ck}(2011)}]{Schollwock2011DMRG}
\bibinfo{author}{\bibfnamefont{U.}~\bibnamefont{Schollw枚ck}},
  \bibinfo{journal}{Annals of Physics} \textbf{\bibinfo{volume}{326}},
  \bibinfo{pages}{96} (\bibinfo{year}{2011}).

\bibitem[{\citenamefont{Sachdev and Keimer}(2011)}]{SachdevPhysToday}
\bibinfo{author}{\bibfnamefont{S.}~\bibnamefont{Sachdev}} \bibnamefont{and}
  \bibinfo{author}{\bibfnamefont{B.}~\bibnamefont{Keimer}},
  \bibinfo{journal}{Physics Today} \textbf{\bibinfo{volume}{64}},
  \bibinfo{pages}{29} (\bibinfo{year}{2011}).

\bibitem[{\citenamefont{Zhang et~al.}(2012)\citenamefont{Zhang, Grover, Turner,
  Oshikawa, and Vishwanath}}]{Zhang2012EE}
\bibinfo{author}{\bibfnamefont{Y.}~\bibnamefont{Zhang}},
  \bibinfo{author}{\bibfnamefont{T.}~\bibnamefont{Grover}},
  \bibinfo{author}{\bibfnamefont{A.}~\bibnamefont{Turner}},
  \bibinfo{author}{\bibfnamefont{M.}~\bibnamefont{Oshikawa}}, \bibnamefont{and}
  \bibinfo{author}{\bibfnamefont{A.}~\bibnamefont{Vishwanath}},
  \bibinfo{journal}{Phys. Rev. B} \textbf{\bibinfo{volume}{85}},
  \bibinfo{pages}{235151} (\bibinfo{year}{2012}).

\bibitem[{\citenamefont{Jiang et~al.}(2010)\citenamefont{Jiang, Rachel, Weng,
  Zhang, and Wang}}]{Jiang2010S2Chain}
\bibinfo{author}{\bibfnamefont{H.-C.} \bibnamefont{Jiang}},
  \bibinfo{author}{\bibfnamefont{S.}~\bibnamefont{Rachel}},
  \bibinfo{author}{\bibfnamefont{Z.-Y.} \bibnamefont{Weng}},
  \bibinfo{author}{\bibfnamefont{S.-C.} \bibnamefont{Zhang}}, \bibnamefont{and}
  \bibinfo{author}{\bibfnamefont{Z.}~\bibnamefont{Wang}},
  \bibinfo{journal}{Phys. Rev. B} \textbf{\bibinfo{volume}{82}},
  \bibinfo{pages}{220403} (\bibinfo{year}{2010}).

\bibitem[{\citenamefont{Stoudenmire et~al.}(2011)\citenamefont{Stoudenmire,
  Alicea, Starykh, and Fisher}}]{Stoudenmire2011Majorana}
\bibinfo{author}{\bibfnamefont{E.~M.} \bibnamefont{Stoudenmire}},
  \bibinfo{author}{\bibfnamefont{J.}~\bibnamefont{Alicea}},
  \bibinfo{author}{\bibfnamefont{O.~A.} \bibnamefont{Starykh}},
  \bibnamefont{and} \bibinfo{author}{\bibfnamefont{M.~P.}
  \bibnamefont{Fisher}}, \bibinfo{journal}{Phys. Rev. B}
  \textbf{\bibinfo{volume}{84}}, \bibinfo{pages}{014503}
  (\bibinfo{year}{2011}).

\bibitem[{\citenamefont{Jiang et~al.}(2008)\citenamefont{Jiang, Weng, and
  Sheng}}]{Jiang2008Kagome}
\bibinfo{author}{\bibfnamefont{H.~C.} \bibnamefont{Jiang}},
  \bibinfo{author}{\bibfnamefont{Z.~Y.} \bibnamefont{Weng}}, \bibnamefont{and}
  \bibinfo{author}{\bibfnamefont{D.~N.} \bibnamefont{Sheng}},
  \bibinfo{journal}{Phys. Rev. Lett.} \textbf{\bibinfo{volume}{101}},
  \bibinfo{pages}{117203} (\bibinfo{year}{2008}).

\bibitem[{\citenamefont{Jiang et~al.}(2009)\citenamefont{Jiang, Kr\"uger,
  Moore, Sheng, Zaanen, and Weng}}]{Jiang2009S1}
\bibinfo{author}{\bibfnamefont{H.~C.} \bibnamefont{Jiang}},
  \bibinfo{author}{\bibfnamefont{F.}~\bibnamefont{Kr\"uger}},
  \bibinfo{author}{\bibfnamefont{J.~E.} \bibnamefont{Moore}},
  \bibinfo{author}{\bibfnamefont{D.~N.} \bibnamefont{Sheng}},
  \bibinfo{author}{\bibfnamefont{J.}~\bibnamefont{Zaanen}}, \bibnamefont{and}
  \bibinfo{author}{\bibfnamefont{Z.~Y.} \bibnamefont{Weng}},
  \bibinfo{journal}{Phys. Rev. B} \textbf{\bibinfo{volume}{79}},
  \bibinfo{pages}{174409} (\bibinfo{year}{2009}).

\bibitem[{\citenamefont{Stoudenmire and White}(2012)}]{Stoudenmire2011}
\bibinfo{author}{\bibfnamefont{E.~M.} \bibnamefont{Stoudenmire}}
  \bibnamefont{and} \bibinfo{author}{\bibfnamefont{S.~R.} \bibnamefont{White}},
  \bibinfo{journal}{Annu. Rev. Condens. Matter Phys.}
  \textbf{\bibinfo{volume}{3}}, \bibinfo{pages}{111} (\bibinfo{year}{2012}).

\bibitem[{\citenamefont{Yan et~al.}(2011)\citenamefont{Yan, Huse, and
  White}}]{White2011Kagome}
\bibinfo{author}{\bibfnamefont{S.}~\bibnamefont{Yan}},
  \bibinfo{author}{\bibfnamefont{D.}~\bibnamefont{Huse}}, \bibnamefont{and}
  \bibinfo{author}{\bibfnamefont{S.}~\bibnamefont{White}},
  \bibinfo{journal}{Science} \textbf{\bibinfo{volume}{332}},
  \bibinfo{pages}{1173} (\bibinfo{year}{2011}).

\bibitem[{\citenamefont{Jiang et~al.}(2012{\natexlab{a}})\citenamefont{Jiang,
  Yao, and Balents}}]{Jiang2011SJ1J2}
\bibinfo{author}{\bibfnamefont{H.~C.} \bibnamefont{Jiang}},
  \bibinfo{author}{\bibfnamefont{H.}~\bibnamefont{Yao}}, \bibnamefont{and}
  \bibinfo{author}{\bibfnamefont{L.}~\bibnamefont{Balents}},
  \bibinfo{journal}{Phys. Rev. B} \textbf{\bibinfo{volume}{86}},
  \bibinfo{pages}{024424} (\bibinfo{year}{2012}{\natexlab{a}}).

\bibitem[{\citenamefont{Depenbrock et~al.}(2012)\citenamefont{Depenbrock,
  McCulloch, and Schollw\"ock}}]{Depenbrock2012Kagome}
\bibinfo{author}{\bibfnamefont{S.}~\bibnamefont{Depenbrock}},
  \bibinfo{author}{\bibfnamefont{I.~P.} \bibnamefont{McCulloch}},
  \bibnamefont{and}
  \bibinfo{author}{\bibfnamefont{U.}~\bibnamefont{Schollw\"ock}},
  \bibinfo{journal}{Phys. Rev. Lett.} \textbf{\bibinfo{volume}{109}},
  \bibinfo{pages}{067201} (\bibinfo{year}{2012}).

\bibitem[{\citenamefont{Jiang et~al.}(2012{\natexlab{b}})\citenamefont{Jiang,
  Wang, and Balents}}]{Jiang2012TEE}
\bibinfo{author}{\bibfnamefont{H.~C.} \bibnamefont{Jiang}},
  \bibinfo{author}{\bibfnamefont{Z.}~\bibnamefont{Wang}}, \bibnamefont{and}
  \bibinfo{author}{\bibfnamefont{L.}~\bibnamefont{Balents}},
  \bibinfo{journal}{Nature Physics} \textbf{\bibinfo{volume}{8}},
  \bibinfo{pages}{902} (\bibinfo{year}{2012}{\natexlab{b}}).

\bibitem[{\citenamefont{Cincio and Vidal}(2013)}]{Cincio2013TO}
\bibinfo{author}{\bibfnamefont{L.}~\bibnamefont{Cincio}} \bibnamefont{and}
  \bibinfo{author}{\bibfnamefont{G.}~\bibnamefont{Vidal}},
  \bibinfo{journal}{Phys. Rev. Lett.} \textbf{\bibinfo{volume}{110}},
  \bibinfo{pages}{067208} (\bibinfo{year}{2013}).

\bibitem[{\citenamefont{Jiang et~al.}(2013)\citenamefont{Jiang, Singh, and
  Balents}}]{Jiang2013TEEAccuracy}
\bibinfo{author}{\bibfnamefont{H.-C.} \bibnamefont{Jiang}},
  \bibinfo{author}{\bibfnamefont{R.~R.~P.} \bibnamefont{Singh}},
  \bibnamefont{and} \bibinfo{author}{\bibfnamefont{L.}~\bibnamefont{Balents}},
  \bibinfo{journal}{Phys. Rev. Lett.} \textbf{\bibinfo{volume}{111}},
  \bibinfo{pages}{107205} (\bibinfo{year}{2013}).

\bibitem[{\citenamefont{Kitaev}(2003)}]{Kitaev2003}
\bibinfo{author}{\bibfnamefont{A.~Y.} \bibnamefont{Kitaev}},
  \bibinfo{journal}{Ann. Phys. (N.Y.)} \textbf{\bibinfo{volume}{303}},
  \bibinfo{pages}{2} (\bibinfo{year}{2003}).

\bibitem[{\citenamefont{Trebst et~al.}(2007)\citenamefont{Trebst, Werner,
  Troyer, Shtengel, and Nayak}}]{Trebst2007}
\bibinfo{author}{\bibfnamefont{S.}~\bibnamefont{Trebst}},
  \bibinfo{author}{\bibfnamefont{P.}~\bibnamefont{Werner}},
  \bibinfo{author}{\bibfnamefont{M.}~\bibnamefont{Troyer}},
  \bibinfo{author}{\bibfnamefont{K.}~\bibnamefont{Shtengel}}, \bibnamefont{and}
  \bibinfo{author}{\bibfnamefont{C.}~\bibnamefont{Nayak}},
  \bibinfo{journal}{Phys. Rev. Lett.} \textbf{\bibinfo{volume}{98}},
  \bibinfo{pages}{070602} (\bibinfo{year}{2007}).

\bibitem[{\citenamefont{Vidal et~al.}(2009)\citenamefont{Vidal, Dusuel, and
  Schmidt}}]{Vidal2008}
\bibinfo{author}{\bibfnamefont{J.}~\bibnamefont{Vidal}},
  \bibinfo{author}{\bibfnamefont{S.}~\bibnamefont{Dusuel}}, \bibnamefont{and}
  \bibinfo{author}{\bibfnamefont{K.~P.} \bibnamefont{Schmidt}},
  \bibinfo{journal}{Phys. Rev. B} \textbf{\bibinfo{volume}{79}},
  \bibinfo{pages}{033109} (\bibinfo{year}{2009}).

\bibitem[{\citenamefont{Tupitsyn et~al.}(2010)\citenamefont{Tupitsyn, Kitaev,
  Prokof'ev, and Stamp}}]{Tupitsyn2008}
\bibinfo{author}{\bibfnamefont{I.~S.} \bibnamefont{Tupitsyn}},
  \bibinfo{author}{\bibfnamefont{A.}~\bibnamefont{Kitaev}},
  \bibinfo{author}{\bibfnamefont{N.~V.} \bibnamefont{Prokof'ev}},
  \bibnamefont{and} \bibinfo{author}{\bibfnamefont{P.~C.~E.}
  \bibnamefont{Stamp}}, \bibinfo{journal}{Phys. Rev. B}
  \textbf{\bibinfo{volume}{82}}, \bibinfo{pages}{085114}
  (\bibinfo{year}{2010}).

\bibitem[{\citenamefont{Gehlen and Rittenberg}(1986)}]{Gehlen1986Q3PottsModel}
\bibinfo{author}{\bibfnamefont{G.~v.} \bibnamefont{Gehlen}} \bibnamefont{and}
  \bibinfo{author}{\bibfnamefont{V.}~\bibnamefont{Rittenberg}},
  \bibinfo{journal}{J. Phys. A: Math. Gen.} \textbf{\bibinfo{volume}{19}},
  \bibinfo{pages}{L625} (\bibinfo{year}{1986}).

\bibitem[{\citenamefont{Feng et~al.}(2007)\citenamefont{Feng, Zhang, and
  Xiang}}]{Feng2007KitaevChain}
\bibinfo{author}{\bibfnamefont{X.-Y.} \bibnamefont{Feng}},
  \bibinfo{author}{\bibfnamefont{G.-M.} \bibnamefont{Zhang}}, \bibnamefont{and}
  \bibinfo{author}{\bibfnamefont{T.}~\bibnamefont{Xiang}},
  \bibinfo{journal}{Phys. Rev. Lett.} \textbf{\bibinfo{volume}{98}},
  \bibinfo{pages}{087204} (\bibinfo{year}{2007}).

\end{thebibliography}

\end{document}